\documentclass[aps,twocolumn,nofootinbib,superscriptaddress,amsfonts,floatfix
]{revtex4-1} 
\usepackage{graphicx,amsmath,amssymb,amstext}
\usepackage{amssymb,amsbsy,amsfonts,amsthm,color}

\usepackage{mathtools,nccmath}

\usepackage{epsfig}
\usepackage{graphicx}
\usepackage{subfigure}
\usepackage[dvipsnames]{xcolor}
\definecolor{linkcolor}{rgb}{0.0,0.3,0.5}
\usepackage[unicode, colorlinks=true, linkcolor=linkcolor, citecolor=linkcolor, 
filecolor=linkcolor,urlcolor=linkcolor, pdfusetitle]{hyperref}
\usepackage{cancel}
\usepackage{balance}
\usepackage[capitalise]{cleveref}
\usepackage{xspace}
\usepackage{enumerate}
\usepackage{ulem}
\usepackage{mdframed}
\usepackage{booktabs}

\usepackage{epigraph}
\AtBeginDocument{%
  \heavyrulewidth=.08em
  \lightrulewidth=.05em
  \cmidrulewidth=.03em
  \belowrulesep=.65ex
  \belowbottomsep=0pt
  \aboverulesep=.4ex
  \abovetopsep=0pt
  \cmidrulesep=\doublerulesep
  \cmidrulekern=.5em
  \defaultaddspace=.5em
}

\graphicspath{{Figures/}}

\usepackage{color}

\newcommand{\Beq}{\begin{eqnarray}}
\newcommand{\Eeq}{\end{eqnarray}}

\def\lsim{\mathrel {\vcenter {\baselineskip 0pt \kern 0pt \hbox{$<$} \kern 0pt \hbox{$\sim$} }}}

\def\gsim{\mathrel {\vcenter {\baselineskip 0pt \kern 0pt \hbox{$>$} \kern 0pt \hbox{$\sim$} }}}

\newcommand{\RomanNumeralCaps}[1]

\def\-{\,-\,}
\def\={\,=\,}
\def\+{\,+\,}
\def\equi{\,\equiv\,}

\definecolor{mypurple}{RGB}{143, 116, 210}

\newcommand{\jhu}{Department of Physics and Astronomy,
Johns Hopkins University, Baltimore, MD 21218, USA}

\begin{document}
{\hfill 
\title{Imaging Topological Solitons: the Microstructure Behind the Shadow}

\author{Pierre Heidmann}
\email{pheidma1@jhu.edu}
\affiliation{\jhu}

\author{Ibrahima Bah}
\email{iboubah@jhu.edu}
\affiliation{\jhu}

\author{Emanuele Berti}
\email{berti@jhu.edu}
\affiliation{\jhu}

\begin{abstract}
 
We study photon geodesics in topological solitons that have the same asymptotic properties as Schwarzschild black holes. These are coherent states in string theory corresponding to pure deformations of spacetime through the dynamics of compact extra dimensions.  We compare these solutions with Schwarzschild black holes by computing null geodesics, deriving Lyapunov exponents, and imaging their geometries as seen by a distant observer.  We show that topological solitons are remarkably similar to black holes in apparent size and scattering properties, while being smooth and horizonless.  Incoming photons experience very high redshift, inducing phenomenological horizon-like behaviors from the point of view of photon scattering. Thus, they provide a compelling case for real-world gravitational solitons and topological alternatives to black holes from string theory. 

\end{abstract}
\maketitle


\section{Introduction} \label{sect:intro2}

The Event Horizon Telescope (EHT) has opened a new observational window on the environment near black holes~\cite{EventHorizonTelescope:2019dse,*EventHorizonTelescope:2022wkp}. Its success and future experimental prospects in imaging could lead to paradigm-changing research in gravitational phenomenology by providing novel ways to explore the strong-gravity environment near black holes, and, as excitingly, the possibility to observe ``exotic'' compact objects beyond general relativity (GR). 

There is a wide variety of interesting proposals for beyond-GR objects and phenomena, including boson stars \cite{PhysRev.172.1331,*Ruffini:1969qy}, gravastars \cite{Mazur:2004fk}, firewalls \cite{Almheiri:2012rt}, non-local interactions \cite{Giddings:2006sj}, soft hair around the horizon \cite{Hawking:2016msc}, and fuzzballs \cite{Mathur:2005zp}. Describing their gravitational signature and their observational differences will offer a promising route for new tests of gravity through direct observations.

In quantum gravity, black holes correspond to thermodynamic ensembles of quantum states.  The general paradigm necessary to fully characterize such states is still lacking.  Often, however, quantum states can be coherent enough to admit classical descriptions.  Indeed, many examples of such states can be constructed from string theory and characterized in various theories of gravity.  Their existence has led to some of the most exciting results in theoretical physics in the last 30 years, such as holography and AdS/CFT \cite{Lin:2004nb,*Kanitscheider:2006zf,*Giusto:2004ip} and black hole microstate geometries \cite{Bena:2022ldq,*Bena:2022rna,*Warner:2019jll}.  The latter objects appear in the fuzzball proposal, which aims to resolve the black hole information paradox in string theory \cite{Mathur:2005zp}.  

The recent developments in EHT and gravitational-wave observations raise the prospect of observing individual coherent states of gravity.  These are necessarily smooth, horizonless and ultra-compact geometries, produced by pure deformations of spacetime without ordinary matter and supported by electromagnetic flux.  For many years, however, these states could only be obtained from supersymmetric theories of gravity, and thereby outside the regime of what we might expect to be astrophysically realistic (see \cite{Bacchini:2021fig,Bena:2020yii,Bena:2020see,*Bianchi:2020bxa,*Bah:2021jno,*Ikeda:2021uvc,Mayerson:2020tpn} for some analysis of their gravitational signatures).  Until recently, it was not clear whether these coherent states can be constructed beyond the lamppost of supersymmetry, and it was widely believed that they could not exist. 

With the motivation of potential observations, two of the authors developed a new framework for constructing solutions in generic, non-supersymmetric, classical theories of gravity with extra compact dimensions that are smooth and horizonless.  The solutions admit gravitational solitons induced by non-trivial topology in the internal space that is supported by electromagnetic flux~\cite{Bah:2020ogh,Bah:2022yji,Bah:2020pdz,*Bah:2021owp,*Bena:2022tro,Heidmann:2021cms,*Bah:2022pdn,*Heidmann:2022zyd}.  Moreover, these solutions can be embedded in string theory, where they can be appropriately interpreted as coherent states of quantum gravity~\cite{Heidmann:2021cms,*Bah:2022pdn,*Heidmann:2022zyd}.  They are characterized and distinguished by their topological microstructure, which necessitates extra compact dimensions.  These are referred to as \emph{topological solitons}.  

In four dimensions, topological solitons manifest as singular ultra-compact objects that are indistinguishable from black holes from afar.  Their higher-dimensional nature becomes apparent at small distances from the solitons, which then resolve the system to smooth and horizonless geometries.  

There exist simple building blocks that can be used to construct complex bound states of topological solitons that may be astrophysically relevant.  These blocks can be seen as a new topological phase of matter that is inherently geometric.  The basic unit corresponds to a spherically symmetric spacetime with a ``bubble of nothing" corresponding to a non-contractible two-cycle.  It is stabilized by adding electromagnetic flux which assigns charges to the bubble \cite{Bah:2021irr,*Stotyn:2011tv,*Miyamoto:2006nd}. This basic ingredient has been named \emph{the topological star} \cite{Bah:2020ogh}.  

By considering bound states of topological stars, we can obtain net-neutral topological solitons with properties comparable to Schwarzschild black holes~\cite{Bah:2022yji}.  These are the first smooth horizonless geometries from string theory that correspond to a Schwarzschild solution far away without any matter in the interior, and they are aptly called \emph{Schwarzschild topological solitons.}  

In this paper, we are interested in describing the phenomenology and gravitational signature of these topological solitons. This will highlight their possible relevance to describe convincing observational alternatives to black holes. As a first step,  we derive their imaging phenomenology and scattering properties by analyzing the behavior of null geodesics in these two types of topological solitons. 

In contrast to gravitational objects produced by ordinary baryonic matter, topological solitons are pure deformations of spacetime with no clear matter delimitation. However, like black holes, we show that these solitons possess an unstable outer photon shell that circumscribes the geometry. Therefore, in the same way that a shadow gives a size to a black hole, we define the \emph{apparent size} of topological solitons by their outer photon shell, as seen by an asymptotic observer. 

Furthermore, photons that enter the shadow of a black hole are absorbed and disappear. The physics is different for smooth and horizonless geometries: incoming photons are non-trivially scattered by the topological microstructure, but re-emerge at some point. From the point of view of an asymptotic observer, the dynamics of these photons carries information about a possible smooth topological microstructure beyond the would-be shadow. We detail the key mechanisms for which such microstructure can still present black hole-like features and induce a phenomenological horizon behavior for photon scattering (these mechanisms have been initially analyzed in Ref.~\cite{Bacchini:2021fig}). The scattering properties of topological solitons in the vicinity of their outer photon shell will thus provide insight into the common features with black holes in GR, but will also identify small deviations that could be useful smoking guns for future imaging experiments.

More precisely, we analyze the properties of the outer unstable photon shells of topological stars and Schwarzschild topological solitons by deriving their apparent size and associated Lyapunov exponents. We also show that the solitons have inner stable photon shells that are not accessible for photons coming from outside the solitons. Moreover, we have built our own ray-tracing code to study their overall gravitational lensing properties, as they would be perceived by a distant observer.

We show that Schwarzschild topological solitons have properties remarkably close to those of Schwarzschild black holes, while being smooth and horizonless.  Their apparent size is strikingly close to the Schwarzschild shadow. Their Lyapunov exponents, redshift, and the time elapsed along the geodesics of initially incoming photons are also remarkably similar.  Overall, they provide the first relevant smooth, horizonless string-theoretic alternatives to non-extremal black holes in GR. 

The paper is organized as follows. In Sec.~\ref{section:inital_data} we review the properties of topological stars and Schwarzschild topological solitons.  In Sec.~\ref{sec:Geo} we analytically derive and analyze their photon shells and some geodesics trajectories. In Sec.~\ref{sec:Imaging} we present our numerical imaging results. In Sec.~\ref{sec:conclusion} we summarize our conclusions and possible directions for future work.

\section{The Geometries}\label{section:inital_data} 

We consider classes of solutions in five-dimensional or six-dimensional Einstein-Maxwell theory with a generic action of the form \cite{Bah:2020ogh,Bah:2022yji}
\begin{equation}
S_D\,=\,\frac{1}{16 \pi G_D} \int d^D x \sqrt{-\det g}\left(R-\frac{1}{2}\left|F\right|^2\right)\,.
\label{eq:Action}
\end{equation}
The solutions are asymptotic to $\mathbb{R}^{1,3} \times$S$^1$ or $\mathbb{R}^{1,3} \times$T$^2$, a product of four-dimensional Minkowski with a circle or a torus.  We parametrize these extra compact dimensions as $y_1$ and $y_2$, and consider that they have a finite and small size asymptotically.  The solutions admit non-trivial topology supported by electromagnetic flux (since we are only interested in uncharged null geodesics in these backgrounds,  we will not specify the flux; we refer the interested reader to Refs.~\cite{Bah:2020ogh,Bah:2022yji} for more details).
They correspond to pure states of gravity that are induced by the dynamics of the extra compact dimensions.  Moreover,  they can be embedded in string theory and admit a description in terms of bound states of strings and branes \cite{Heidmann:2021cms,*Bah:2022pdn,*Heidmann:2022zyd}.
Note that the charges under the gauge fields can be interpreted as ``hidden dark charges,'' such that they only interact gravitationally with ordinary baryonic particles and are not ruled out by current observations \cite{Cardoso:2016olt,*Bozzola:2020mjx,Mayerson:2020tpn,Bacchini:2021fig}. 

\subsection{The topological star}
\label{sec:TS}

Topological stars are five-dimensional horizonless geometries that are spherically symmetric and static.  They can be labeled  by two parameters $r_\text{B} > r_\text{S}\geq 0$ and they have a metric
\begin{align}
ds_5^2 \,=\, &-\left( 1- \frac{r_\text{S}}{r}\right) \,dt^2+\frac{dr^2}{\left( 1- \frac{r_\text{S}}{r}\right) \left( 1- \frac{r_\text{B}}{r}\right) } + r^2 \,d\Omega_2^2 \nonumber\\
&+ \left( 1- \frac{r_\text{B}}{r}\right) \,dy_1^2 \,,
\label{eq:metTS}
\end{align}
where $d\Omega_2^2= d\theta^2+\sin^2 \theta \,d\phi^2$ is the line element of a two-sphere.  The solutions carry a magnetic charge \cite{Bah:2020ogh}
\begin{equation}
Q\,=\, \sqrt{3r_\text{S}r_\text{B}}\,,
\label{eq:chargeTS}
\end{equation}
where we have considered that the electric coupling is $e=(16\pi G_5)^{-1/2}$.
The spacetime is smooth and terminates at $r=r_\text{B}$ where the $y_1$-circle smoothly collapses,  defining the origin of a $\mathbb{R}^{2}$ space.  In this region, the geometry has an $\mathbb{R}^2 \times$S$^2$ topology and thus admits a S$^2$ bubble with radius $r_\text{B}$. Regularity imposes an algebraic constraint between $r_\text{B}$, $r_\text{S}$ and the extra-dimension size, $R_{y_1}$. However, adding a conical defect at the bubble allows $r_\text{B}$ and $r_\text{S}$ to decouple from  $R_{y_1}$ \cite{Bah:2020ogh}. We depict a typical geometry in Fig.~\ref{fig:SchematicTS}.

\begin{figure}
\begin{center}
\includegraphics[width=0.4\textwidth]{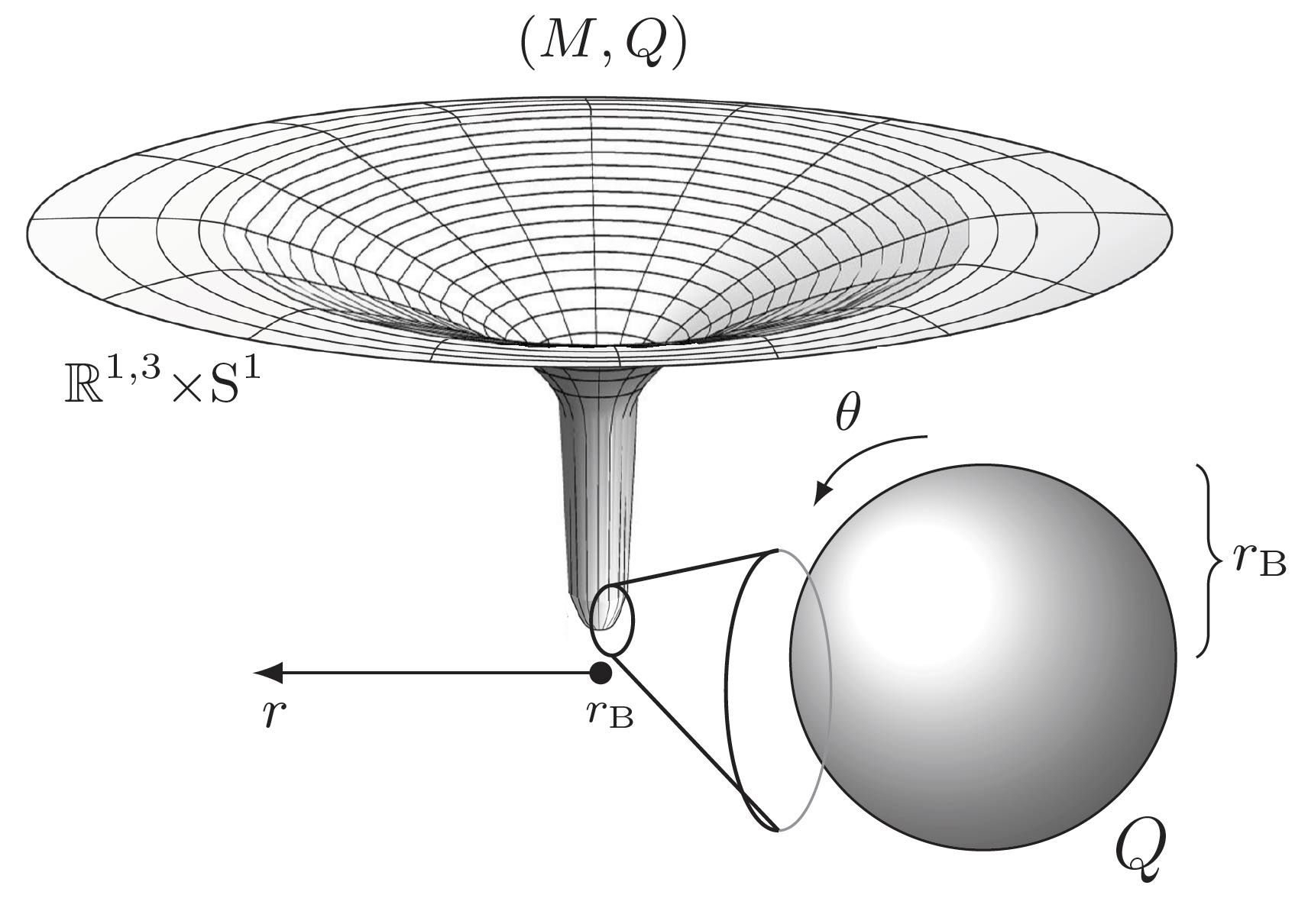}
\caption{Schematic description of a topological star.  The spacetime is smooth and terminates at $r=r_\text{B}$ where the $y_1$ circle degenerates.  This induces a large topological bubble of charge $Q$ that can be considered as the ``surface'' of the star.}
\label{fig:SchematicTS}
\end{center}
\end{figure}

Upon compactifying on the $y_1$-circle, topological stars can be described by singular four-dimensional solutions with ADM mass (in units where $G_4=1$)
\begin{equation}
M \,=\, \frac{2 r_\text{S} + r_\text{B}}{4}\,.
\end{equation}
The solutions correspond to classically and thermodynamically meta-stable states of the theory of gravity \eqref{eq:Action} if and only if \cite{Bah:2021irr,*Stotyn:2011tv,*Miyamoto:2006nd}
\begin{equation}
r_\text{S}\,<\, r_\text{B} \,<\,  2\,r_\text{S}\,,
\label{eq:stabilityrange}
\end{equation}
which we assume from now on.  Topological stars admit an extremal limit when $r_\text{S}\to r_\text{B}$,  where they approach extremal black string solutions~\cite{Bah:2020ogh}.

Finally,  it is also useful to parametrize the solitons in terms of asymptotic quantities measurable at infinity: the mass and the charges.  In the range \eqref{eq:stabilityrange},  there exists one topological star at given mass and charge:
\begin{equation}
r_\text{S} \= M+\sqrt{M^2-\tfrac{1}{6}Q^2}\,,\quad r_\text{B} \= 2\left(M-\sqrt{M^2-\tfrac{1}{6}Q^2}\right).
\label{eq:rsrbinversion}
\end{equation}
We consider solutions in the range
\begin{equation}
\frac{4}{\sqrt{3}} < \mu <\sqrt{6}\,,\qquad \mu \equi \frac{Q}{M}\,.
\label{eq:mass&chargeRange}
\end{equation}
The extremality bound is at $\mu \sim 4/\sqrt{3}$,  while the stability bound is at $\mu \sim \sqrt{6}$. 

As objects labeled by mass and charge, the domain of validity of the topological stars differs from Reissner-Nordstr\"om black holes (which exist for $\mu \leq 1$) and from the five-dimensional black string in this same theory (which exist for $\mu \leq \frac{4}{\sqrt{3}}$). In this paper, we will study geodesics of topological stars and compare them to Schwarzschild black holes of the same mass.

\subsection{The Schwarzschild topological soliton}
\label{sec:STS}

The Schwarzschild topological solitons constructed in Ref.~\cite{Bah:2022yji} are six-dimensional smooth horizonless solutions that are axially symmetric and static. They correspond to net-neutral bound states of three topological stars in six dimensions.  They consist of a  chain of three bubbles where the $y_1$ and $y_2$ circles smoothly degenerate.  The two outermost bubbles carry opposite charges,  while the one in the middle is uncharged.  We have depicted the profile of the geometry in Fig.~\ref{fig:SchematicSTS}.  

In contrast to the single topological star, the present soliton is {\em neutral} while being supported by electromagnetic flux, and is therefore in the same regime of mass and charge as a Schwarzschild black hole.  The solutions are given in terms of two mass parameters,  $\ell$ and $m$,  and a parameter $q$ related to the amplitude of the internal charges.  Since we are dealing with bound states of three bubbles,  we introduce three local spherical coordinates centered around each bubble:
\begin{figure}[t]
\begin{center}
\includegraphics[width=0.43\textwidth]{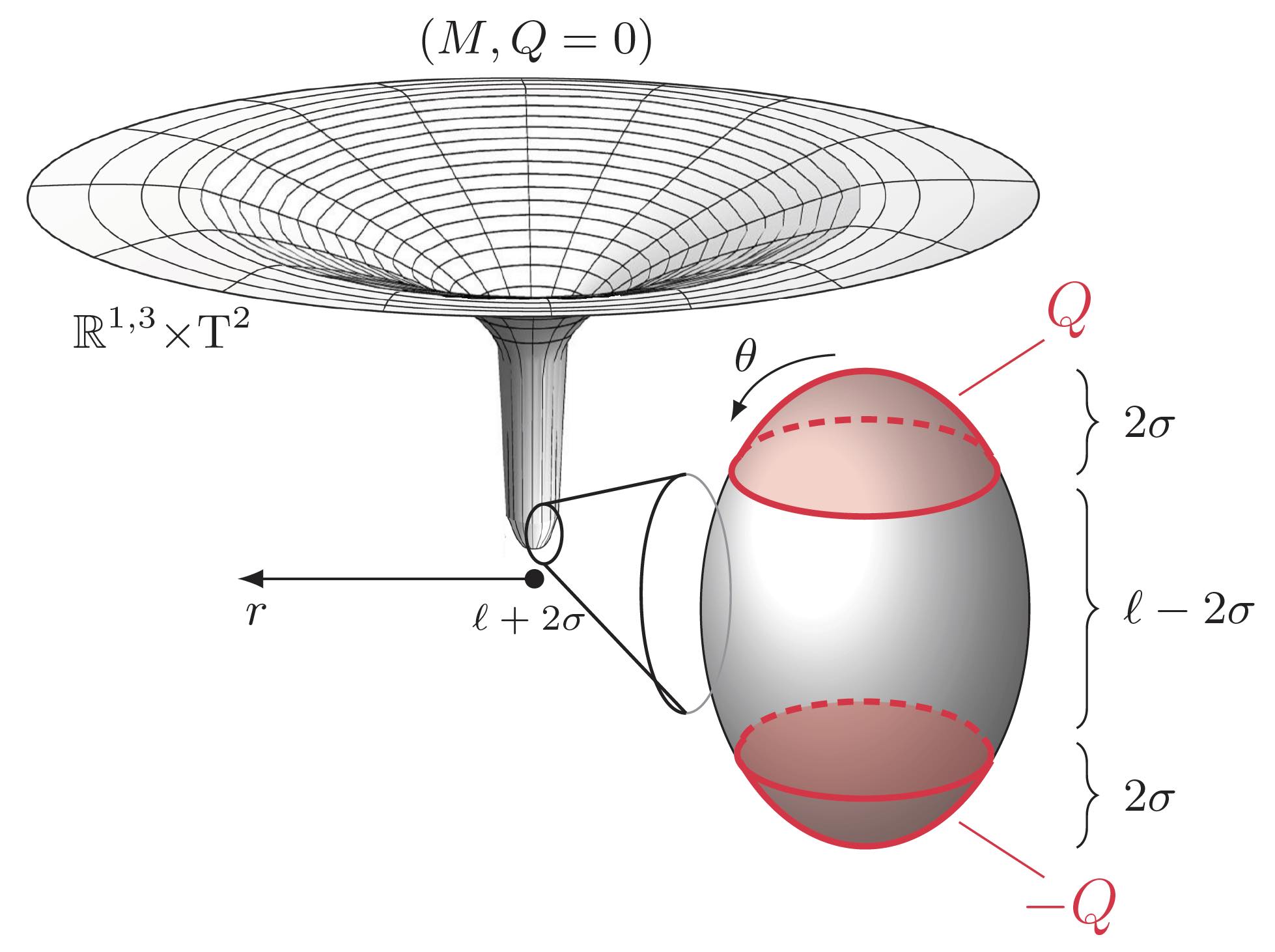}
\caption{Schematic description of a Schwarzschild topological soliton.  It is a neutral  bound state of three topological stars.  The spacetime is smooth and terminates at $r=\ell+2\sigma$ where the $y_1$ and $y_2$ circles degenerate alternatively.  This induces a large topological bubble that can be considered as the ``surface'' of the soliton.}
\label{fig:SchematicSTS}
\end{center}
\end{figure}
\begin{align}
 r_1 &\,\equiv\, \frac{r_-^{(0)} +r_-^{(1)}}{4}- \sigma\,, \quad \cos \theta_1 \,\equiv\,  \frac{r_-^{(0)} -r_-^{(1)}}{4\sigma}\,,\nonumber \\
 r_2 &\,\equiv\, \frac{r_-^{(1)} +r_+^{(1)}}{4}-\frac{1}{2}(\ell-2 \sigma)\,, \quad \cos \theta_2 \,\equiv\,  \frac{r_-^{(1)} -r_+^{(1)}}{2(\ell-2 \sigma)}\,,\nonumber \\
 r_3 &\,\equiv\, \frac{r_+^{(1)} +r_+^{(0)}}{4}- \sigma\,, \quad \cos \theta_3 \,\equiv\,  \frac{r_+^{(1)} -r_+^{(0)}}{4\sigma}\,, 
\end{align}
where we have defined $2\sigma$ to be the size of both outermost bubbles, and four distances, $(r_\pm^{(0)},r_\pm^{(1)})$,  that depend on the main spherical coordinates $(r,\theta)$ as follows:
\begin{align}
\sigma & \,\equiv \, \sqrt{m^2-q(q-2\gamma)}\,,\qquad \gamma \,\equiv \, \frac{2mq}{\ell+2m}\,,\nonumber \\
r_\pm^{(0)}& \equiv 2 r - (\ell+2\sigma)(1\pm \cos \theta)\,,  \\
r_\pm^{(1)}  &\equiv\medmath{\sqrt{((2(r-\sigma)-\ell)\cos \theta\pm(2\sigma-\ell))^2+4 r (r-\ell-2\sigma)\sin^2\theta}} .\nonumber
\end{align}
Smooth and regular solutions exist when the parameters satisfy the conditions~\cite{Bah:2022yji}:
\begin{equation}
\ell > 2 m \,,\qquad q < m \,\sqrt{\frac{\ell+2m}{\ell-2m}}.
\end{equation}
The second inequality corresponds to the extremal bounds of the two outer bubbles, where they degenerate into extremal black holes.
\vspace{0.5cm}
\begin{widetext}
The metric is given by
\begin{align}
ds_6^2 \,=\, &\frac{1}{Z} \left[ -dt^2 + \frac{r_1 r_3}{(r_1+2\sigma)(r_3+2\sigma)}\,dy_1^2\right] + \frac{r_2\,Z}{r_2+\ell-2\sigma}\, dy_2^2+ Z \left[f \left( \frac{r \,dr^2}{r-\ell-2\sigma} +r^2 \,d\theta^2\right) + r^2 \sin^2\theta \,d\phi^2 \right]\,,
\label{eq:metSTS}
\end{align}
where we have introduced the functions
\begin{align}
Z\,\equiv \, &\frac{(r_1+\sigma + m)(r_3+\sigma +m)+\left(q-\gamma(1+\cos \theta_3)\right)\left(q-\gamma(1-\cos \theta_1)\right)}{\sqrt{\left((r_1+2\sigma)^2+\gamma^2\,\sin^2 \theta_3\,\left(1+\frac{2\sigma}{r_1}\right)\right)\left((r_3+2\sigma)^2+\gamma^2\,\sin^2 \theta_1\,\left(1+\frac{2\sigma}{r_3}\right)\right)}}, \nonumber \\
f^2 \,\equiv \, & \frac{1}{(1+2\delta)^2}\,\frac{(r_1(r_1+2\sigma)+\gamma^2\sin^2\theta_3)\,\,(r_3(r_3+2\sigma)+\gamma^2\sin^2\theta_1)}{\left(\left(r_2+\tfrac{1}{2}(\ell-2\sigma) \right)^2- \tfrac{1}{4}(\ell-2\sigma)^2\cos^2 \theta_2\right) \,\,\left(r_1+\sigma(1-\cos \theta_1) \right)\,\,\left(r_3+\sigma(1+\cos \theta_3) \right)} \label{eq:mainfunctionSTS} \\ 
&\times \frac{r_1(1+\cos\theta_1)+r_3(1-\cos\theta_3)}{r_1(1-\cos\theta_1)+r_3(1+\cos\theta_3)} \,\,  \frac{(r_1+2\sigma)(1-\cos\theta_1)+(r_3+2\sigma)(1+\cos\theta_3)}{(r_1+2\sigma)(1+\cos\theta_1)+(r_3+2\sigma)(1-\cos\theta_3)}\nonumber \\
& \times \left(1+2\delta \,\frac{(q-\gamma)(r_1-r_3)+(\gamma m-\ell(q-\gamma))(\cos \theta_1+\cos \theta_3)}{(q-\gamma)(r_3-r_1)+\gamma m (\cos \theta_1+\cos \theta_3)}\right)^2\,,\qquad\delta \,\equiv \, \frac{m^2(\ell+2m)^2+\ell^2 q^2}{(\ell+2m)^2(\ell^2-2m^2)+2\ell^2 q^2}\,.\nonumber
\end{align}

\end{widetext}
The spacetime is smooth and terminates at $r=\ell+2\sigma$. At this locus, either $r_1=0$,  $r_2=0$ or $r_3=0$, depending on the value of $\theta$.  The ranges of $\theta$ are delimited by the critical angle $\cos \theta_c = \tfrac{\ell-2\sigma}{\ell+2\sigma}$.  For $0\leq\theta\leq \theta_c$ and $\pi-\theta_c\leq \theta\leq \pi$,  $r_3=0$ and $r_1=0$, respectively. In these regions the $y_1$-circle degenerates.  For $\theta_c\leq \theta\leq \pi-\theta_c$,  $r_2=0$,  and the $y_2$-circle degenerates.  Regularity conditions lead to algebraic constraints on the parameters in terms of extra-dimension sizes \cite{Bah:2022yji}.  In this paper, we restrict to a part of the phase space that has properties similar to Schwarzschild black holes:
\begin{equation}
\begin{split}
\ell &= \frac{4M}{3}\left(1+\epsilon_2\right), \qquad m=\frac{2 M}{3}\left(1-\frac{\epsilon_2}{2}\right), \\
 q&=\frac{4 M}{3 \sqrt{3 \epsilon_2}}(1-\epsilon_1),
\end{split}
\label{eq:ParamSTS}
\end{equation}
where $M$ is the four-dimensional ADM mass of the soliton
\begin{equation}
M \,= \, \frac{4m+\ell-2\sigma}{4}\,,
\end{equation}
and $(\epsilon_1,\epsilon_2)$ are infinitesimal parameters related to the ratios of the extra-dimension sizes with the ADM mass.  In this limit, the two outer charged bubbles are very close to their extremal limit $\sigma =\mathcal{O}(\epsilon_1 M)$.

Away from the bubbling structure, the solutions are approximated with extreme precision by a singular vacuum solution given by the four-dimensional metric (see Ref. \cite{Bah:2022yji} for more details)
\begin{align}
d s_4^2=&-\left(1-\frac{4M}{3 r}\right)^{\frac{3}{2}} d t^2+r^2\left(1-\frac{4 M}{3 r}\right)^{-\frac{1}{2}} \sin ^2 \theta \,d \phi^2 \nonumber \\
&+\frac{\left(1-\frac{4 M}{3 r}\right)^{\frac{1}{2}}\left(d r^2+r^2\left(1-\frac{4M}{3 r}\right) d \theta^2\right) }{\left[\left(1-\frac{2 M}{3 r}\right)^2-\left(\frac{2 M}{3 r}\right)^2 \cos ^2 \theta\right]^2}.
\label{eq:ApproxGeo}
\end{align}
This axially-symmetric solution has a naked singularity at $r=4M/3$.  Our solitons are indistinguishable up to a scale infinitesimally close to this locus, and resolve the singularity into a smooth bound state of bubbles in six dimensions.  

\noindent The approximated singular solution is not geodesically complete at $r=4M/3\sim \ell+2\sigma$.  At this locus,  one must consider the full bubbling solutions given in Eq.~\eqref{eq:metSTS}, for which the spacetime terminates smoothly there.  However,  the approximated geometry is helpful to describe the dynamics of null geodesics for trajectories that do not get too close to the soliton surface.

\section{Null geodesics}\label{sec:Geo}

We aim at describing the physics of null geodesics in the classes of topological solitons introduced above.  They are generically given by the equations
\begin{align}
&\ddot{x}^\mu  \+ \Gamma^\mu_{\,\,\alpha \beta}\,\dot{x}^\alpha\,\dot{x}^\beta\=0\,, \quad x^\mu\equiv(t,r,\theta,\phi,y_1,y_2),  \label{eq:GeoEqGen}
\end{align}
where $\Gamma^\mu_{\,\,\alpha \beta}$ is the Levi-Civita connection and we have defined $\dot{x}=\frac{dx}{d\tau}$,  so that $\tau$ is an affine parameter.  Null geodesics in backgrounds with four isometries along $(t,\phi,y_1,y_2)$ have five constants of motion: the Hamiltonian $H=g_{\mu \nu} \,\dot{x}^\mu \dot{x}^\nu=0$ and the momenta associated to the four commuting Killing vectors $p_\mu\equi g_{\mu \nu} \,\dot{x}^\nu$.  This leads to
\begin{align}
&\dot{t} \= - g^{tt}\,,\quad \dot{\phi} \= g^{\phi \phi} \,p_\phi\,,\quad \dot{y}_a \=  g^{y_a y_a} \,p_{y_a}\,,\quad a=1,2\,,\nonumber \\
&g_{rr}  \dot{r}^2 +g_{\theta \theta} \dot{\theta}^2 \= -g^{tt}-g^{\phi\phi} \,p_\phi^2 -g^{y_1 y_1} p_{y_1}^2-g^{y_2 y_2} p_{y_2}^2\,,
\label{eq:GeoPotGen}
\end{align}
where the dependence on $y_2$ must be dropped for the topological stars.

The momenta along the extra dimensions induce an effective mass  since $g^{y_a y_a} p_{y_a}^2 \to p_{y_a}^2$ at large distance $r\to\infty$. The only difference with massive probes is that this effective mass term increases as the probe approaches the soliton. This should produce small deviations for massive trajectories that approach it closely. Moreover, $p_{y_a}$ necessarily scales as ``$(\text{extra-dimensional sizes})^{-1}$,'' and a probe with momentum corresponds to a very massive and highly excited particle that is unlikely to be produced in any physical process. Therefore, we will focus on massless geodesics from a four-dimensional perspective. This requires $p_{y_1}=p_{y_2}=0$.  

\subsection{Photon scattering in topological stars}
\label{sec:AnalyticTS}

Topological stars are spherically symmetric backgrounds, so we limit attention to geodesics in the equatorial plane without loss of generality.  The equations reduce to the following radial equation \cite{Lim:2021ejg,Guo:2022nto}:
\begin{equation}
\dot{r}^2- V_\text{TS} \= 0\,, \quad V_\text{TS}  \equiv\left(1-\frac{r_\text{B}}{r}\right) \left[1-\left(1-\frac{r_\text{S}}{r}\right) \frac{p_\phi^2}{r^2}\right]\,.
\end{equation}

\subsubsection{Photon spheres and Lyapunov exponent}

Topological stars have photon spheres,  obtained by imposing $\dot{r}=\ddot{r}=0$.  There are two solutions, labeled by the radii of the photon sphere $(R_1,R_2)$ and given by
\begin{equation}
\begin{split}
R_1 &\= r_\text{B} \= 2\left(M-\sqrt{M^2-\tfrac{1}{6}Q^2}\right)\,,\\
 R_2& \=\frac{3}{2} r_\text{S} \=  \frac{3}{2}\left(M+\sqrt{M^2-\tfrac{1}{6}Q^2}\right) \,,
 \end{split}
 \label{eq:PSTS}
\end{equation}
Their associated angular momenta and angular velocity,  $p_\phi$ and $\Omega=\dot{\phi}/\dot{t}$,  are $p_1 = \Omega_1^{-1}=\frac{r_\text{B}^\frac{3}{2}}{\sqrt{r_\text{B}-r_\text{S}}}$ for the first photon sphere,  and $p_2=\Omega_2^{-1}=\frac{3\sqrt{3}r_\text{S}}{2}$ for the second photon sphere.

An important observation here is that the ``end-to-spacetime'' locus,  i.e.,  the surface of the star $r=r_\text{B}$,  is a photon sphere.  Therefore,  \emph{a photon can be trapped at the surface of the topological star}. This unusual feature is dramatically different from other compact objects.  

The second photon sphere, $R_2$, is not always part of the spacetime if we take into account the stability bound of Eq.~\eqref{eq:stabilityrange}.  Thus,  depending on whether $3 r_\text{S}$ is greater or smaller than $2 r_\text{B}$,  we have one or two photon spheres.  There are \emph{two kinds of topological stars} according to this attribute: the topological stars of the first kind with one photon sphere,  and the topological stars of the second kind with two photon spheres. They exist in different ranges of $(r_\text{S}, r_\text{B})$, which can be translated to different ranges of mass-to-charge ratios, as depicted in Fig.~\ref{fig:Lyap&TSkinds}.

\begin{figure}[t]
\begin{center}
    {\includegraphics[width=\columnwidth]{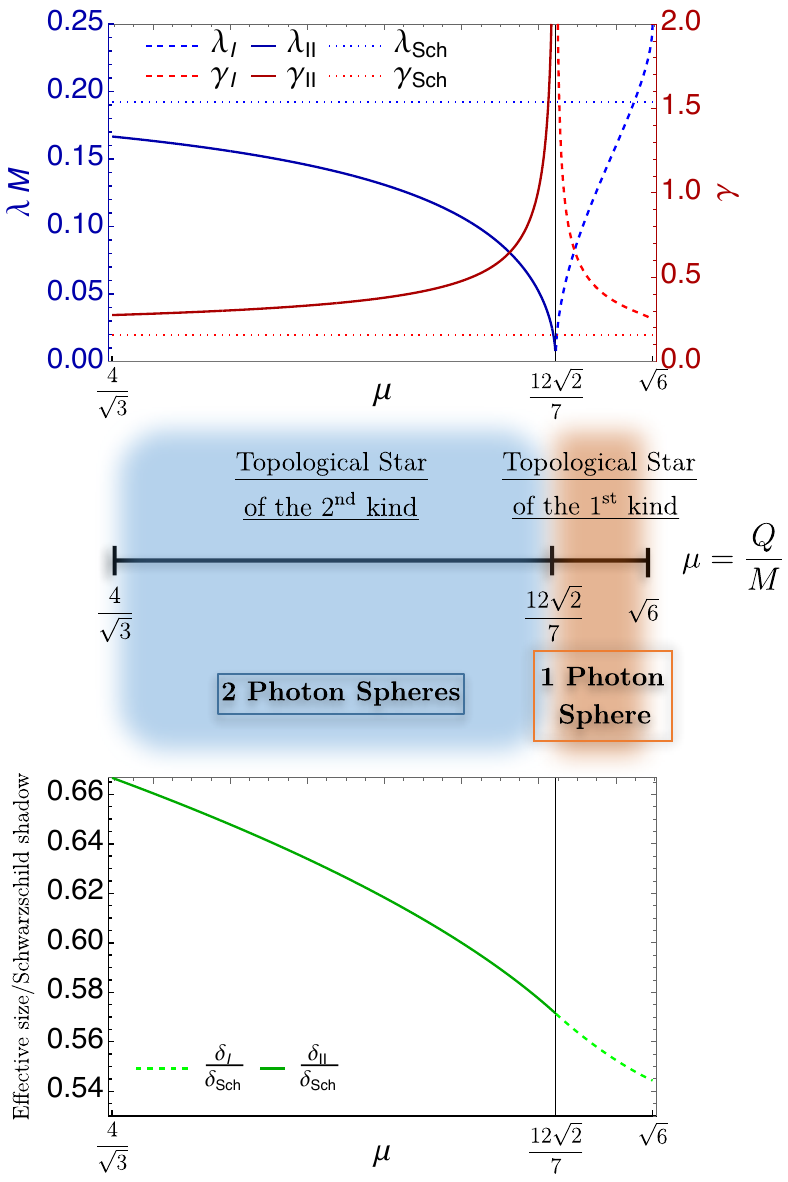}}
\caption{Scattering properties of topological stars as a function of the charge-to-mass ratio $\mu$. Top panel: the Lyapunov and critical exponents associated to their outer unstable photon sphere.  Middle panel: Description of the two kinds of topological stars. Bottom panel: their apparent size with respect to the Schwarzschild shadow, as defined in Sec.~\ref{sec:EffSizeTS}.}
\label{fig:Lyap&TSkinds}
\end{center}
\end{figure}
 
The photon spheres are stable (unstable) if the second derivative of the potential with respect to $r$ is negative (positive).  We find that, for both kinds of topological stars,  \emph{the outermost photon sphere is unstable}.  However,  for the topological stars of the second kind, the inner photon sphere  (the surface of the star) is stable. This highlights long-term trapping effects for such geometries, and leads to an interesting spectrum containing slowly-damped quasinormal modes \cite{QNMsTS}. This long-term trapping has been  analyzed in the context of other topological solitons in string theory that are coherent manifestations of supersymmetric black hole microstates \cite{Eperon:2016cdd,*Marolf:2016nwu,*Bena:2019azk,*Bianchi:2021xpr,*Consoli:2022eey,Bena:2020yii}.  It is a well-understood instability in black hole microstates, that is key to resolve black hole information paradox.
Note however that no geodesics coming from outside the outer photon sphere can be  trapped at the inner sphere.  This is only possible for photons that originate in between $R_1$ and $R_2$.  
 
One can derive Lyapunov exponents associated to the unstable photon spheres.  For spherically symmetric spacetimes they are generically given by \cite{Cornish:2003ig,Cardoso:2008bp}
\begin{equation}
\lambda \= \sqrt{\frac{1}{2 \dot{t}^2}\,\frac{d^2V_\text{TS}}{dr^2}} \Bigl|_{r=R_a}\,.
\end{equation}
We find
\begin{equation}
\lambda_\text{I} =\frac{\sqrt{(r_\text{B}-r_\text{S})(2 r_\text{B}-3 r_\text{S})}}{r_\text{B}^2}\,,\quad \lambda_\text{II} = \frac{2\sqrt{3 r_\text{S}-2 r_\text{B}}}{9 r_\text{S}^\frac{3}{2}} \,,
\end{equation}
where $\lambda_\text{I}$ and $\lambda_\text{II}$ are the Lyapunov exponents of the first and second kinds of topological stars, respectively.  One can also write the exponents in terms of the mass and charge using Eq.~\eqref{eq:rsrbinversion}.  

We have plotted the Lyapunov exponents as a function of $\mu$ and in unit of $M^{-1}$ in Fig.~\ref{fig:Lyap&TSkinds}.  The exponents vary in the range $[0,\,\frac{1}{4}]$.  Even if one cannot compare rigorously to Schwarzschild black holes since we are dealing with charged geometries, the maximum is larger than the value of the Lyapunov exponent at the Schwarzschild shadow, $\lambda_\text{Sch} = 1/(3\sqrt{3}) M^{-1} \approx 0.19 M^{-1}$.   

Therefore topological stars have properties similar to black holes: they induce strong gravitational lensing, and they have a photon sphere surrounding the geometry that is highly sensitive to initial boundary conditions.   Note that this does not necessarily mean that geodesics in topological stars have chaotic behavior~\cite{Cornish:2003ig}.   Indeed,  the geodesic equations are integrable, and the Lyapunov exponents at the photon spheres just indicate the average rate of expansion or contraction of adjacent geodesics in the phase space.

Finally,  following Pretorius and Khurana~\cite{Pretorius:2007jn},  we compute the critical exponent measuring the ratio between the Lyapunov instability time scale with the orbital time scale, given by
\begin{equation}
\gamma_c \= \frac{\dot{\phi}}{2\pi\lambda \,\dot{t}}.
\end{equation}
We find that
\begin{align}
\gamma_{c\text{I}} = \frac{1}{2\pi}\,\sqrt{\frac{r_\text{B}}{2r_\text{B}-3 r_\text{S}}},\quad \gamma_{c\text{II}} = \frac{1}{2\pi}\,\sqrt{\frac{3r_\text{S}}{3 r_\text{S}-2r_\text{B}}}.
\end{align}
These critical exponents are also plotted in the top panel of Fig.~\ref{fig:Lyap&TSkinds}. They are always greater than the Schwarzschild exponent $\gamma_\text{Sch}=(2\pi)^{-1}$ and they diverge as $Q/M \to 12\sqrt{2}/7$,  which is when the stable and unstable photon spheres merge.

As argued in Ref.~\cite{Pretorius:2007jn}, the critical exponent $\gamma_c$ is relevant for critical phenomena in binary mergers, in particularly to understand whether only a fraction of the total energy of the system can be radiated in the ultrarelativistic limit~\cite{Sperhake:2008ga}.

In the encounter of two black holes, three
outcomes are possible. For large values of the impact parameter of the collision, the black holes just scatter off each other. At small values of the impact parameter, they merge directly following a nearly radial plunge. However there is also an intermediate
regime leading to a {\em delayed merger}, where the black holes can revolve around each other (in principle) an infinite number of times by fine-tuning the impact
parameter around some critical value $b=b_*$.

Reference~\cite{Pretorius:2007jn} observed that $\gamma_c$ is proportional to the number of orbits spent in this critical region and conjectured that in the ultrarelativistic limit, by fine-tuning $b$ around $b_*$, the two black holes could in principle radiate all of their kinetic energy. The critical exponent plays a critical role in this conjecture, at least in the extreme mass-ratio limit \cite{Berti:2010ce}. 

An important observation is in order: while for black holes we can directly relate the Lyapunov exponents to the black hole's quasi-normal modes~\cite{Cardoso:2008bp}, this is not possible for topological solitons.  Indeed,  this computation implicitly requires the presence of a horizon at which the scalar waves satisfy ingoing boundary conditions.  The physics is different in the context of smooth horizonless geometries, as we will discuss in a separate study~\cite{QNMsTS}.

\subsubsection{Apparent size of topological stars}
\label{sec:EffSizeTS}

Topological stars,  and more generically smooth topological structures in string theory,  are not made of ordinary matter but correspond to coherent states of gravity.  There is no stress-energy tensor that can be used to define traditional features such as their size.  Their phenomenological attributes are inherent to the spacetime itself, in analogy with horizons for black holes.  The geometric size of topological solitons is given by various topological cycles at the locus where the spacetime ends.  

Another way to associate a size to compact object such as black holes is to consider their shadow, i.e., their photon shell. Topological stars have an outer photon sphere that surrounds the object and gives them an apparent size. Unlike black holes,  this photon sphere does not delimit a sharp shadows, since geodesics that go in also come out.  The photon sphere is more appropriately seen as a region for extreme gravitational lensing.  This is similar to the effect of black hole shadows, but with small differences that we will highlight in Sec.~\ref{sec:Imaging}.

Using a simple parallax relation,  we define the apparent size $\delta$ of a topological soliton as the size of its outer photon shell as seen by an asymptotic observer:
\begin{equation}
\delta \= \underset{r\to \infty}{\lim} \frac{r^2 \dot{\phi}}{\dot{r}},
\label{eq:ApparentSizeGen}
\end{equation}
where $\dot{\phi}$ is fixed such that $p_\phi$ is the critical momentum associated to the outer photon sphere.  For the Schwarzschild metric we have the well-known result
\begin{equation}
\delta_\text{Shadow} \= 3\sqrt{3} M \approx 5.2 M\,,
\label{eq:ApparentSizeSch}
\end{equation}
while for the two kinds of topological stars we find 
\begin{align}
\delta_\text{I}& \= \frac{r_\text{B}^\frac{3}{2}}{\sqrt{r_\text{B}-r_\text{S}}} \= \frac{2\sqrt{2} \left( 1-\sqrt{1-\tfrac{1}{6}\mu^2}\right)^{3/2}}{3\sqrt{3} \sqrt{1-3\sqrt{1-\tfrac{1}{6}\mu^2}}}\, \delta_\text{Shadow}\,.\nonumber \\
\delta_\text{II}& \= \frac{3\sqrt{3}r_\text{S}}{2} \=\frac{1+\sqrt{1-\tfrac{1}{6}\mu^2}}{2}\,  \delta_\text{Shadow} \,. \label{eq:ApparentSizeTS}
\end{align}
We have plotted these apparent sizes as a function of the charge-to-mass ratio in the bottom panel of Fig.~\ref{fig:Lyap&TSkinds}.  Topological stars of the first kind are slightly more compact than topological stars of the second kind.  Moreover,  topological stars always look more compact than a Schwarzschild black hole of the same mass: the aspect ratios range from $0.66$ to $0.54$.  This is perhaps not surprising: increasing the charge increases the energy density, and thereby should lead to stronger gravitational attraction.

\subsection{Photon scattering in Schwarzschild solitons}
\label{sec:AnalyticSTS}

Schwarzschild topological solitons are axially symmetric solutions, and thus the geodesic equations are not integrable. Computing photon trajectories is therefore a challenge.  The upside is that this implies a greater diversity of  trajectories and more novel features as compared to spherically symmetric systems.   

 Some techniques have been developed in the context of four-dimensional black hole bound states (see, e.g., Ref.~\cite{Liang:1974ha,*Shipley:2016omi,*Dolan:2016bxj,Cunha:2018acu}).  However,  the aim of this paper is not to have an exhaustive analytic classification of null geodesics.  We will consider only a few illustrative examples that allow us to obtain an apparent size given by the outermost photon shell of the solitons,  and its associated Lyapunov and critical exponents.  

Null geodesics are given by a two-dimensional potential
\begin{equation}
\begin{split}
&\dot{r}^2+r^2\left(1-\frac{\ell+2\sigma}{r}\right)\dot{\theta}^2  - V_\text{STS} \=0\,,\\
&V_\text{STS} \= \frac{\left(1-\frac{\ell+2\sigma}{r} \right)}{f} \left(1-\frac{p_\phi^2}{r^2Z^2\sin^2\theta }\right)\,,
\label{eq:PotSTS}
\end{split}
\end{equation}
where $f$ and $Z$ are defined in Eq.~\eqref{eq:mainfunctionSTS}.  Using a numerical approach,  that we will detail in the next section,  we have observed that the solitons have an outermost photon shell at $r \sim 8M/3 \sim 2\ell$.  It has an ellipsoidal shape due to the axial symmetry,  and is slightly flattened at its poles.  Moreover,  it has similar scattering properties to the Schwarzschild shadow:  all trajectories that are slightly outside the photon shell escape the geometries, while incoming photons necessarily reach the end-to-spacetime locus, $r=\ell+2\sigma$,  and remain trapped for a long time.

We will characterize the photon shell numerically in Sec.~\ref{sec:Imaging}.  To obtain some analytic results, we consider the scattering properties of this outermost photon shell and its apparent size by focusing on some trajectories that depend on one variable only.

\subsubsection{Outer photon shell and Lyapunov exponent}

Since the solitons are $\mathbb{Z}_2$-symmetric,  $\theta \to \pi-\theta$,  the two-dimensional potential necessarily has $\partial_\theta V_\text{STS} =0$ at $\theta=\pi/2$.  Therefore we can study geodesics on the equatorial plane.
At $\theta =\frac{\pi}{2}$,  the geodesics are governed by a one-dimensional potential obtained from \eqref{eq:PotSTS}.  We have $V_\text{STS} = \partial_r V_\text{STS}=0$ if and only if
\begin{equation}
p_\phi^2 = r^2 Z^2 |_{\theta=\tfrac{\pi}{2}} \,,\qquad (r-\ell-2\sigma)\, \partial_r( r Z)  |_{\theta=\tfrac{\pi}{2}} =0\,.
\end{equation}
The function $r Z$,  defined in \eqref{eq:mainfunctionSTS}, has a global minimum around $r\sim 2\ell \sim 8M/3$ \eqref{eq:ParamSTS}.  Thus, there are two photon orbits on the equator:  one at the end-to-spacetime locus $r =\ell+2\sigma$,  and another at around twice the distance. The latter corresponds to the outermost photon shell of the solitons,  restricted to the equatorial plane. 
Moreover,  the sign of $\partial_r^2 V_\text{STS}$ indicates that the outer orbit is unstable under perturbation along the radial direction, while the inner one is stable.  This is similar to the topological star of the second kind.

We derive the Lyapunov and critical exponents associated to the outer photon orbit,  which measure the instability in the equatorial plane.  For that purpose,  one can make use of the approximate geometry introduced in Eq.~\eqref{eq:ApproxGeo}.  Indeed,  our solitons are almost indistinguishable from  the metric \eqref{eq:ApproxGeo} around $r\sim 8M/3$,  and will therefore have the same properties (modulo small corrections of the order of the extra-dimension sizes).  The geodesic potential on the equatorial plane of the metric \eqref{eq:ApproxGeo} is given by
\begin{equation}
\begin{split}
V_\text{app} \= &\frac{\left(1-\frac{2M}{3r} \right)^4}{\left(1-\frac{4M}{3r} \right)^2}  \left[ 1- \left(1-\frac{4M}{3r} \right)^2 \frac{p_\phi^2}{r^2} \right]\,.
\end{split}
\end{equation}
This potential has indeed a photon orbit at $r=\frac{8M}{3}$ satisfying
\begin{equation}
\lambda_\text{STS} \= \frac{27}{128 M} \,,\qquad \gamma_{c\text{STS}} \= \frac{4}{9\pi} \,.
\end{equation}
Its associated angular momentum and angular velocity are $p_\phi=\Omega^{-1}=\frac{16 M}{3}$, respectively. Therefore, the two  exponents and the angular velocity associated to the unstable equatorial orbit are remarkably close to the Schwarzschild values,  since we have $( \gamma_{c\text{STS}}, \gamma_{c\text{Sch}}) \approx (0.16,0.14)$, $(\lambda_\text{STS},\lambda_\text{Sch})\approx (0.21,0.19)\,M^{-1}$ and $(\Omega,\Omega_\text{Sch})\approx (0.192,0.187)\,M^{-1}$.  If the angular velocities are almost identical,  the Lyapunov exponent of the soliton is slightly larger. Therefore,  the scattering properties of our neutral smooth horizonless geometries are remarkably close to Schwarzschild, but more unstable in the vicinity of its outermost photon orbit.

\subsubsection{Apparent size of Schwarzschild topological solitons}

As in the case of topological stars,  the photon shell that surrounds the geometry gives an apparent size to the soliton.  On the equator, the photon orbit is at $r=8M/3 \lesssim 3M$, which is the radius of the Schwarzschild shadow.  While in the spherically symmetric case of the topological star $r$ is a scalar and defines physical radius of a sphere, this is not the case in the axially-symmetric system.
Nonetheless, the apparent size of the soliton can be defined as in Eq.~\eqref{eq:ApparentSizeGen},  and we find
\begin{equation}
 \delta_\text{STS} \= \frac{16}{3} M \approx 5.3 M\,.
\label{eq:ApparentSizeSTS}
\end{equation}
This value is once again remarkably close to the Schwarzschild apparent size \eqref{eq:ApparentSizeSch}, and so an asymptotic observer will barely notice the difference between the parallax angles.  

One should a priori do a similar computation out of the equatorial plane to obtain the total size of the solitons.  Due to the complexity of the geodesic equations,  this is only possible numerically.  However,  we will see in the next section that the outer photon shell is actually almost spherically symmetric and slightly flattened at its poles, so that the apparent size on the equatorial plane \eqref{eq:ApparentSizeSTS} is a good approximation of the total apparent size of the Schwarzschild topological solitons.

At first sight the metric of a Schwarzschild topological soliton, which is well approximated by Eq.~\eqref{eq:ApproxGeo}, is very different from the Schwarzschild metric.  However,  its outer photon shell is \emph{remarkably similar in size and scattering properties} to the shadow of a Schwarzschild black hole. For such gravitational objects without matter sources,  the photon shell or the shadow define the size and the effective properties of the geometry. We therefore expect that their images will look very similar.  However,  unlike the shadow,  all photons that enter the soliton can come out.  This is the main difference,  and in the rest of the paper we will explore numerically how this difference affects the scattering and imaging properties of these objects.

\section{Imaging}\label{sec:Imaging}

In this section,  we study null geodesic trajectories as seen from a distant observer using numerical ray-tracing methods. This has been done previously for other exotic ultra-compact geometries (see e.g.~\cite{Cunha:2018acu} for a review, and \cite{Bohn:2014xxa,*Cunha:2015yba,*Bacchini:2018zom,*Hertog:2019hfb,Bacchini:2021fig} for a non-exhaustive list of recent work).  We constructed our own code since the ones in the literature require a four-dimensional metric.  Our geometries are well captured by four-dimensional physics up to the soliton surface. However, the extra compact dimensions are needed to resolve the singularity of the effective four-dimensional system in order to have a geodesically complete spacetime.  

We applied our code to four types of geometries of the same mass $M = 3/4$ (this choice ensures that the topological star of the second kind and the Schwarzschild soliton both $r_\text{B}\sim \ell \sim 1$), and also to empty flat space for calibration. The four geometries are a Schwarzschild black hole, the two kinds of topological stars (with charge taken in their range of validity in Fig.~\ref{fig:Lyap&TSkinds}), and the  Schwarzschild topological soliton.  

Our main results are presented in Fig.~\ref{fig:Fig1}.  Moreover, since the Schwarzschild topological soliton is not spherically symmetric, in Fig.~\ref{fig:Fig2} we analyze its imaging properties for different inclination angles of the observer with respect to the axis of symmetry.

\subsection{Methodology}

\begin{figure*}[t]
\begin{center}
    {\includegraphics[width=2\columnwidth]{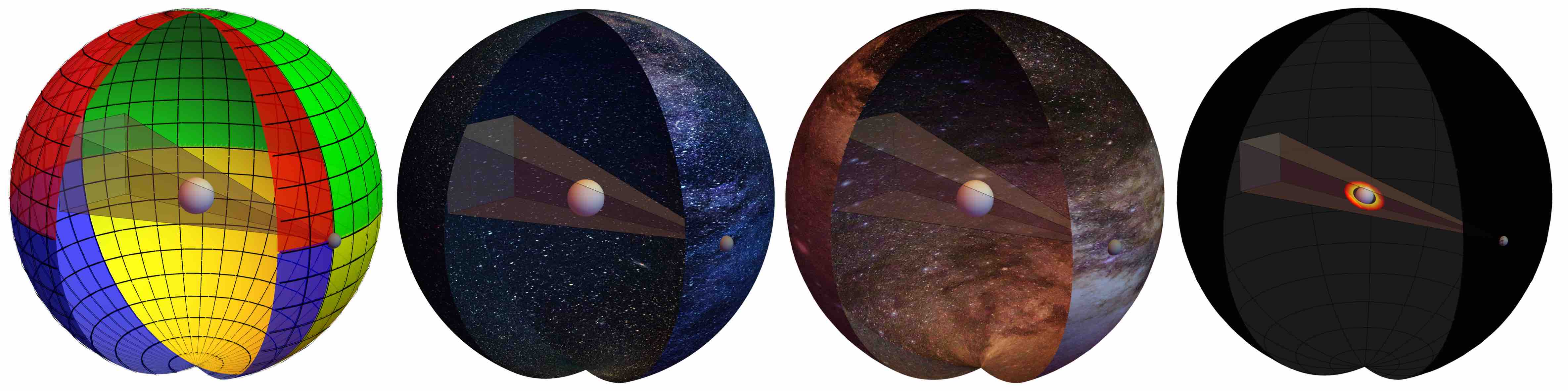}}
\caption{Illustration of the artificial background grids.  The camera (gray point) is on a ``celestial'' sphere centered around the gravitational objects (represented as a white ball).  The sphere has a large radius with respect to the object size: $R=20 M$ for the three first backgrounds,  and $R=40M$ for the last.  For the first background,  the sphere is covered by a quadri-color patch with a grid of meridians and latitudes of angle $\pi/20$.  For the second and third backgrounds,  the celestial sphere is covered by pictures of the Milky Way.  For the last background,  the sphere is totally black,  and there is a bright ``accretion disk.''  The disk has a $\pi/3$ inclination angle with respect to the camera-object plane and  the disk radius is in the range $[(3+1/3)M,\,5M]$.  The celestial spheres have been artificially cut here to improve the readability of the figures.
    }
\label{fig:Methodo}
\end{center}
\end{figure*}

We solve numerically the geodesic equations \eqref{eq:GeoEqGen} and impose the null condition, $g_{\mu\nu} \dot{x}^\mu \dot{x}^\nu=0$,  on the initial data.  To avoid any issues with the geodesics approaching the degeneracy locii of the solitons,  we solve the equations in terms of the proper radial coordinate,  given by $\rho^2=r-r_\text{B}$ for the topological stars and $\rho^2=r-\ell-2\sigma$ for the Schwarzschild topological soliton.  Once the geodesic equations are solved,  we switch back to the $r$ coordinate.

The observer is placed on a ``celestial'' sphere at a large radius $R$ where the spacetime is mostly flat (see Fig.~\ref{fig:Methodo}). We choose $R=20 M$ in most of the cases, and $R=40M$ in one situation that we specify in Fig.~\ref{fig:Methodo}.  We consider a camera of  $10^6$ pixels that is pointing towards the center of the spacetime, with an angle of view of $\delta \varphi = 2\pi/7$ when $R=20M$ and $\delta \varphi = \pi/7$ when $R=40M$.

The geodesics are numerically integrated backwards in time from the camera to where they originated on the celestial sphere.  More precisely we shoot $10^6$ geodesics with different angles $(\varphi_1,\varphi_2)$ in the camera frame, and integrate them until they cross the sphere.  To highlight the gravitational effects of the backgrounds on the null geodesics, we track the following physical quantities:
\begin{itemize}
  \setlength\itemsep{-0.2em}
\item[•] Their original position on the celestial sphere,  given by $(\theta,\phi)$. Note that the original position corresponds to the ending position from the point of view of the integration.
\item[•] The total time elapsed along the geodesics,  $\Delta t$,  where $t$ is the time measured on the celestial sphere.
\item[•] The maximum redshift,  $ \max (-g^{tt})$,  experienced by the geodesics along the trajectory.
\item[•] The coordinates of the intersection,  if it exists,  between  the geodesics and an effective ``accretion disk'' centered around the geometry. The disk has a $\pi/3$ inclination angle with respect to the plane of the camera and the center of mass,  and its radius is ranging from $(3+1/3)M$ to $5M$.  Note that our disk is chosen for illustrative purposes only, and it lies below the Schwarzschild innermost stable circular orbit bound at $6M$. Our objective is more a conceptual comparison of the solitons and the Schwarzschild metric than a realistic imaging simulation for future EHT experiments, which could be a topic for future projects.  As the solitons are very similar to Schwarzschild, a disk located close to the ``shadow'' is needed to highlight possible deviations.
\end{itemize}

From these quantities,  we construct several illustrative graphs.

We first highlight the distorted apparent sky seen by the camera by dividing the celestial sphere into four quadrants,  each painted with a different color and a grid of meridians and latitudes (leftmost panel in Fig.~\ref{fig:Methodo}).  We assign a definite color to each pair $(\theta,\phi)$ on the celestial sphere,  and reconstruct the picture on the camera by taking their associated $(\varphi_1,\varphi_2)$ after scattering.

Using the same method,  we then produced slightly more ``arty'' pictures by covering the celestial sphere with images of stars in the Milky way (second and third panels in Fig.~\ref{fig:Methodo}).

Then we simulate the picture that could be obtained when a bright accretion disk orbits around the geometries (right panel in Fig.~\ref{fig:Methodo}).  This is a simple illustration of what could be seen by the Event Horizon Telescope in the future \cite{EventHorizonTelescope:2019dse,EventHorizonTelescope:2022wkp,Mayerson:2020tpn,Carballo-Rubio:2022aed}.   To reduce the irrelevant gravitational lensing effect at ``short'' distance, we double the radius of the celestial sphere and divide the camera angle by two.

We plot the elapsed time of the geodesics in units of the mass of the geometry and as a function of the position on the camera $(\varphi_1,\varphi_2)$. This estimates the chaoticity experienced by the geodesics and the failure of the probe approximation.   Indeed, a long elapsed time generally implies 
a greater chance for the photon to be absorbed by the geometry.  

We also plot the maximum redshift experienced by null particles along the trajectory normalized to the redshift at the celestial sphere.  This provides, together with the elapsed time,  a measure of how much energy a geodesic would lose by escaping the soliton \cite{doi:10.1142/4890}.  More concretely,  this allows us to go beyond the probe computation and to estimate a {\em darkening factor} for highly-redshifted trajectories of order $\sqrt{-\max g^{tt}}$~\cite{doi:10.1142/4890,Bacchini:2021fig}.

We apply these steps to a flat spacetime for calibration, and to four typical spacetimes of identical mass $M=3/4$:
\begin{itemize}
  \setlength\itemsep{-0.2em}
\item[•] A topological star of the first kind with $r_\text{B} = 1.4$,  $r_\text{S}=0.8$, and charge $Q \approx 1.83$ [Eq.~\eqref{eq:chargeTS}]. 
\item[•] A topological star of the second kind with $r_\text{B} = 1.04$,  $r_\text{S}=0.98 $, and charge $Q \approx 1.75$ [Eq.~\eqref{eq:chargeTS}]. 
\item[•] A Schwarzschild topological soliton with $\ell=1.005$, $m=0.49875$ and $q=8.1496$.  The parameters are chosen so that we are in the Schwarzschild regime given by Eq.~\eqref{eq:ParamSTS}.  
\item[•] A Schwarzschild black hole of mass $M=3/4$.
\end{itemize}

\subsection{Results and analysis}

Our results are presented in Figs.~\ref{fig:Fig1} and \ref{fig:Fig2},  which we analyze here.

\begin{figure*}[t]
\begin{center}
    {\includegraphics[width=2.1\columnwidth]{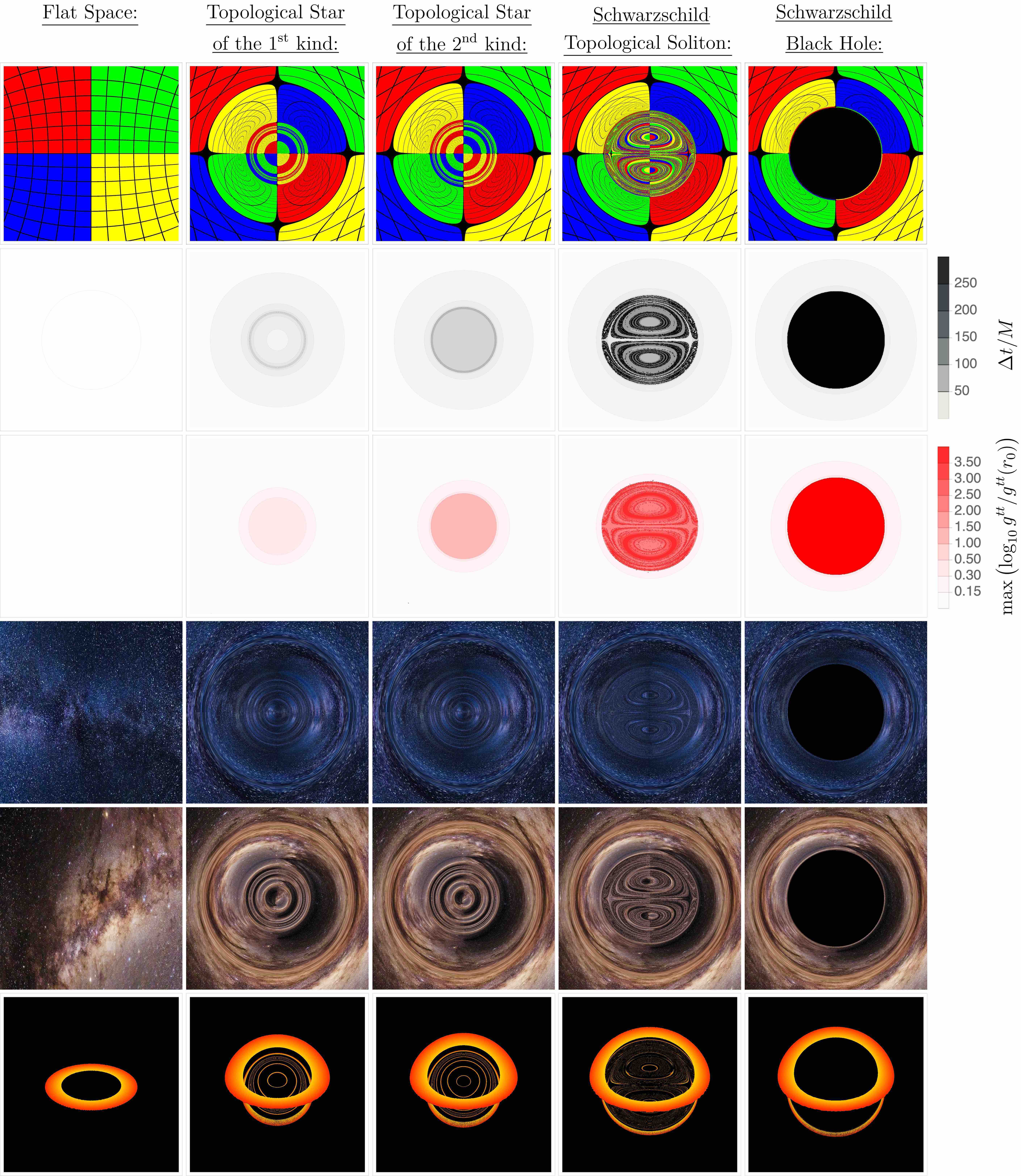}}
\caption{Gravitational lensing effects of the smooth horizonless solitons compared to flat space and a Schwarzschild black hole.  From left to right: the five different backgrounds,  flat space,  the two kinds of topological stars,  the Schwarzschild topological soliton, and the Schwarzschild black hole.  From top to bottom: the quadri-color screen,  the elapsed time,  the maximum redshift experienced,  the two ``arty'' sky screens,  and the accretion disk picture. }  
\label{fig:Fig1}
\end{center}
\end{figure*}

\begin{figure}[h]
\begin{center}
    {\includegraphics[width=0.92\columnwidth]{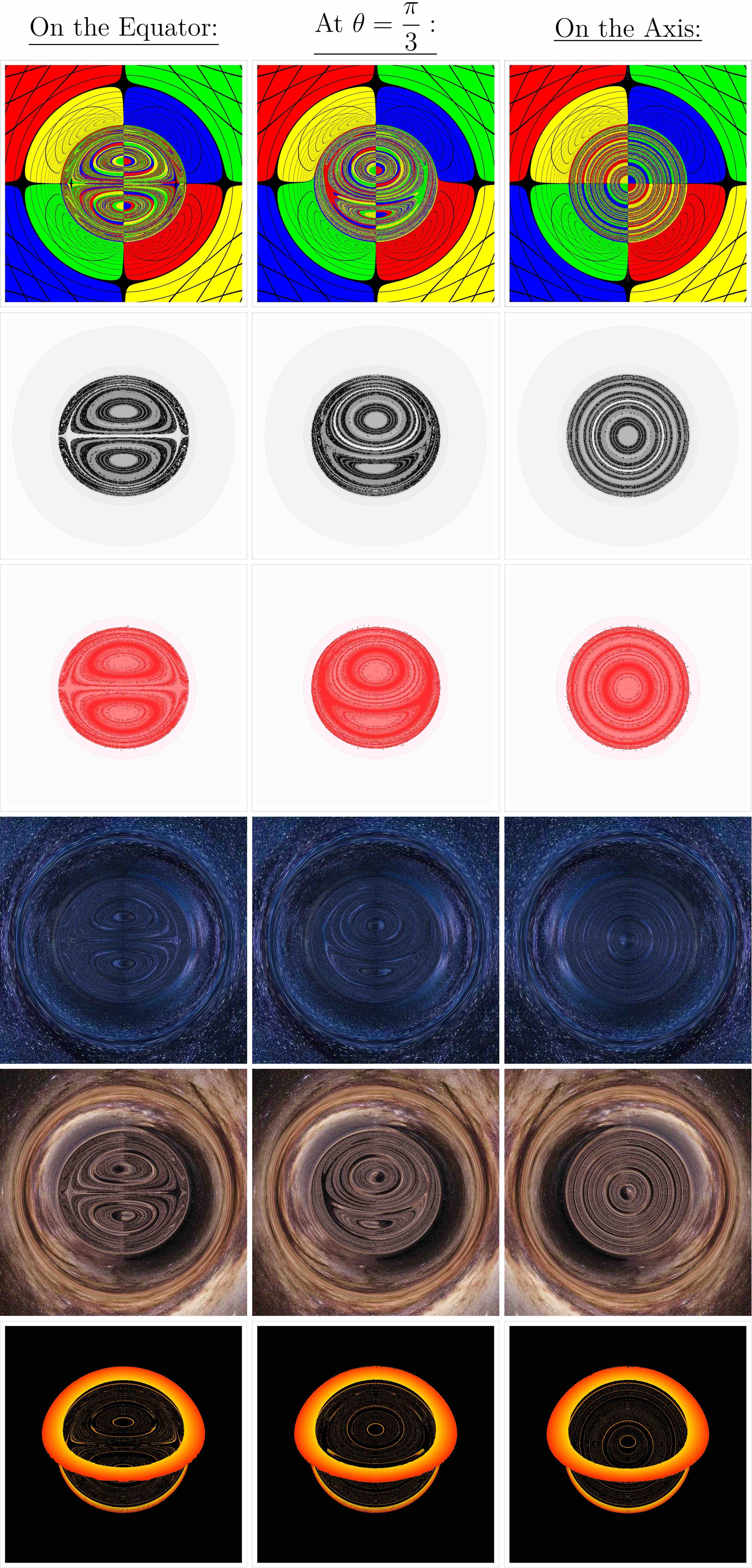}}
\caption{Gravitational lensing effects of the Schwarzschild topological soliton as a function of the angle between the observer and the axis of symmetry.  From left to right: the observer is on the equatorial plane of the soliton $\theta=\pi/2$,  on the North-hemisphere side at $\theta=\pi/3$, and on the axis at $\theta=0$.  The rows follow the same conventions as in Fig.~\ref{fig:Fig1},  and the scale for the elapsed time (second row) and maximum redshift experienced (third row) are also the same. }
\label{fig:Fig2}
\end{center}
\end{figure}

\subsubsection{Gravitational lensing and photon orbits}

We first discuss the lensing effect of the smooth horizonless solitons in comparison to black holes in GR using the quadri-color imaging, and also the more ``arty'' pictures in Figs.~\ref{fig:Fig1} and~\ref{fig:Fig2}. 

As expected, topological stars have a strongly coherent effect on the trajectories due to the integrable structure of their geodesic equations.  From the point of view of an asymptotic observer, they are circumscribed by their outermost unstable photon spheres.  Topological solitons are pure deformations of spacetime, with no ordinary matter radiating out and indicating the surface of the star.  Thus, like black holes, their apparent size is determined by the size of their photon shell as seen from an asymptotic observer: see Eq.~\eqref{eq:ApparentSizeTS}.  However, unlike black holes, the photon sphere is not a shadow: all trajectories that enter bounce back.  For topological stars, the bouncing is very simple, so that when the trajectory reaches the ``end of spacetime'' at $r=r_\text{B}$, it bounces back with mirror symmetry.  Therefore, topological stars behave like \emph {``spherical spacetime mirrors''} for lights.  However, the trajectories are still strongly curved, so that they produce non-trivial rings where photons rotate several times around the solitons.  In each ring,  and especially in the central one,  the whole spacetime gets reflected coherently. This feature produces a ``water wave surface'' effect in the ``arty'' pictures at the bottom of Fig.~\ref{fig:Fig1}.

As argued in Sec.~\ref{sec:AnalyticTS},  the stable inner photon sphere of the second kind of topological stars is not visible by the observer.  Therefore,  both kinds of topological stars are relatively similar from afar.  Moreover,  their apparent sizes are very much comparable, as we analytically derived in Sec.~\ref{sec:EffSizeTS}.  They are also significantly smaller than the Schwarzschild shadow, although the comparison of a charged gravitational object with a neutral one must be taken with a grain of salt.

On the other side,  the imaging of the Schwarzschild topological solitons can be directly compared to the Schwarzschild pictures,  since they are both neutral.  As expected from Sec.~\ref{sec:AnalyticSTS},  the two geometries look remarkably similar in several aspects. The outermost photon shell of the solitons is extremely close to the Schwarzschild shadow.  While the size has been derived analytically in the equatorial plane in Eq.~\eqref{eq:ApparentSizeSTS},  we can see now that it has an ellipsoidal shape that is smaller than the Schwarzschild shadow along the axis of symmetry.

Moreover,  the topological solitons have replaced the inside of the shadow with regular gravitational structures inducing chaotic scattering behaviors that are expected from coherent black hole microstates.  First, the quadri-color screen reveals that increasingly many trajectories traveling inside the would-be black hole shadow follow chaotic paths. They are scattered across the celestial sphere and form fewer coherent structures, as indicated by the larger chaotically-colored regions.  This is a direct consequence of having non-integrable geodesic equations. 

However,  the picture still has some remnants of coherence through multiple internal photon rings, as we can see in Fig.~\ref{fig:Fig1} (and also in Fig.~\ref{fig:Fig2}, when the inclination angle varies).  The Schwarzschild topological soliton considered here is the most primitive neutral soliton supported by electromagnetic flux one can build: the solutions are axially-symmetric bound states of three bubbles with a dipolar structure.   As a consequence,  this dipolar symmetry has an imprint on the scattering properties, such that some geodesics can orbit coherently around each individual bubble and induce the ring patterns we observe.  However,  considering more generic and less symmetric geometries will lead to more chaotic and less coherent gravitational lensing effects. 

The ``arty'' pictures for the Schwarzschild solitons highlight a more realistic consequence of these chaotic features. Unlike topological stars, the geodesics that travel inside the would-be black hole shadow get scattered away as a blurred homogeneous cloud with a faded coherent dipolar pattern.  Therefore, they provide realistic alternatives for exotic smooth regular geometries from quantum gravity beyond black holes in GR.

In conclusion, the probe calculation confirms the analytical results.  Schwarzschild topological solitons have scattering properties very close to those of Schwarzschild black holes. These include their apparent size,  scattering angles, etcetera.  However,  the smooth microstructure inside the soliton reflects the light chaotically, unlike black holes.  We will estimate the fate of these reflections by going one step beyond the probe picture.

\subsubsection{Chaoticity,  redshift and effective scrambling}

The redshift and elapsed-time plots allow us to go beyond the probe description of the scattering properties.  Indeed, when backreaction is included, light with nearly trapped chaotic trajectories and high redshift begins to interact with the background and lose energy.  This produces an effective scrambling behavior, as expected in black holes, but caused here by a physically regular mechanism. The photons in these trajectories, even if they escape within the probe calculation, are expected to be significantly redshifted and fall outside the detectable wavelength range \cite{doi:10.1142/4890,Mayerson:2020tpn,Bacchini:2021fig}.  Therefore, the elapsed time and experienced redshift, combined with chaotic motions, exhibit \emph{horizon-like characteristics} in smooth topological geometries without horizons.

In Fig.~\ref{fig:Fig1},  we show that topological stars are once again very coherent geometries.  They do not induce a long-term trapping to photons that go inside the geometries, except for those concentrated on the photon orbits.  The redshift experienced is relatively small, with a maximum value of order $10$ for topological stars of the second kind.  Therefore,  {generic topological stars do not generate strong black-hole-like effects on scattering photons.

For the Schwarzschild topological solitons,  the photons crossing the outermost photon shell experience extreme redshift and time delay, and therefore have horizon-like properties.  It is also remarkable how the pictures look similar to the scattering in supersymmetric classical fuzzballs~\cite{Bacchini:2021fig}, which are known to be gravitational manifestations of quantum microstates of extremal supersymmetric black holes with very long throats.   

From the second row in Fig.~\ref{fig:Fig2},  we can see that the regions of high elapsed time correspond to the highly chaotic regions in the quadri-color pictures.  They correspond to complicated trajectories that are orbiting back and forth around the bubbles forming the bound states.

As explained in Sec.~\ref{sec:STS},  the Schwarzschild topological solitons consist of a vacuum bubble having near-extremal bubbles with opposite charges at its poles.  The latter have almost zero size, and are smooth topological resolutions of extremal black strings.  Therefore,  the solitons have a very high redshift at the location of the charged bubbles (see Ref.~\cite{Bah:2022yji} for more details on the exact values of the redshift in terms of the mass and extra-dimension sizes).  In the third row of Fig.~\ref{fig:Fig2},  the high-redshift regions correspond to the trajectories that go very close to these bubbles.  Overall,  the redshift experienced by photons that go inside the outer photon shell ranges from 10$^2$ to 10$^4$,  and it can be even higher for solutions where the charged bubbles are even closer to their extremal limit.

Moreover,  the energy loss of the photons can be well approximated by the maximum redshift encountered along the trajectory. In this sense, the redshift is a good approximation for the ``darkness'' that must be added to the probe computation \cite{doi:10.1142/4890}.  According to that property,  we could have therefore darkened the ``arty'' pictures in Fig.~\ref{fig:Fig2} by a factor $\sqrt{\text{max}(-g^{tt})}$ to obtain more realistic photon scattering images. This would make the Schwarzschild topological solitons almost indistinguishable from a Schwarzschild black hole, since all of the patterns inside the photon shell would barely be visible.

Furthermore,  from the EHT measurements~\cite{EventHorizonTelescope:2019dse,EventHorizonTelescope:2022wkp},  the upper bound on the image brightness at the center of M87* is $<10\%$.  By assuming that this implies a $10^{-2}$  redshift on average,  the Schwarzschild topological soliton presented in this analysis is already within this bound.  While it is an interesting example of what may exist according to string theory, it excitingly raises the prospect of horizonless stringy alternative for Schwarzschild black holes in the real world.

\subsubsection{Accretion disk}

Let us now discuss the last row of Figs.~\ref{fig:Fig1} and \ref{fig:Fig2}, where we have done a simple simulation of the image produced by adding a bright accretion disk orbiting around our geometries.

For these plots, we have approximately taken into account the effect of the redshift on the expected luminosity of the scattered photons. More specifically, we have reduced the luminosity by $10\%$ for each unit of $\max \log_{10} (-g^{tt})$ (for example, the luminosity of a photon with maximum redshift of $10^4$ is reduced by $40\%$).    This is a much smaller darkening factor than we might realistically expect, since we should simply multiply the luminosity by $\sqrt{\text{max}(-g^{tt})}$.   However,  the approximation gives a good visual idea of the darkening effect, while still allowing scattered geodesics to be discerned in the high-redshift region of the geometries.

For a Schwarzschild black hole, the light coming from the front of the disk goes directly to the observer, so it is almost undistorted.  In contrast, the light coming from the other side is strongly bent.  Because of this lensing, a significant part of the light is seen at a higher inclination and a small ``tail'' appears from below, outlining the shadow of the black hole~\cite{Luminet:1979nyg}.  

We observe the same generic properties for the solitons due to their outer photon shell that surrounds the geometries.  However,  as these are smooth topological geometries, some of the photons are reflected in a non-trivial way by passing inside the topological solitons and bouncing off.  For topological stars, the reflections are very coherent circles, as if the light had been reflected by a spherical ``spacetime mirror." For the Schwarzschild topological solitons, most of the light that goes in is chaotically scattered, forming a residual glow from inside the soliton. Some trajectories however still follow the coherent dipolar patterns formed by the bound states, inducing coherent ring reflections from the inside.

Remarkably,  the image of the Schwarzschild topological soliton is once again almost indistinguishable from the image of a Schwarzschild black hole. This similarity would have been even more striking if we had darkened the pictures by a more realistic factor of $\sqrt{\text{max}(-g^{tt})}$. The main difference are the ``inner reflections'' in the solitons.  This is a remarkable feature since the metrics are very different, even outside the Schwarschild horizon.  However,  since their outermost photon shell is very much comparable in size and in terms  of Lyapunov exponents, the two classes of geometries have very similar gravitational lensing properties.  Being able to differentiate them with EHT will require a great improvement in resolution to detect more precisely what is scattered from inside the would-be black hole shadow \cite{Carballo-Rubio:2022aed}.  

The Schwarzschild topological solitons are therefore the first compelling non-supersymmetric and non-extremal geometries that are manifestations of quantum gravity states of matter, and that have scattering properties very similar to Schwarzschild.  They are the first bound states of strings and branes in string theory that demonstrate the existence of viable alternative to astrophysical black holes in GR.

\section{Conclusions and outlook}\label{sec:conclusion}

In this paper, we have analyzed the properties of four-dimensional string-theoretic horizonless topological solitons that are non-extremal and can have macroscopic properties comparable to astrophysical black holes.  These new objects constructed by two of the authors provide a compelling case for gravitational solitons in the real world from string theory.

We considered two types of geometries. The first, topological stars, are simple spherically-symmetric geometries that have a non-zero dark charge~\cite{Bah:2020ogh}. The second, Schwarzschild topological solitons, are neutral axially-symmetric bound states of topological stars that can have the same mass and charge as four-dimensional Schwarzschild black holes~\cite{Bah:2022yji}.

We focused on the physics of null geodesics and light scattering in these backgrounds, highlighting differences and similarities with ordinary black hole solutions in GR.  Topological solitons are coherent states of gravity that emerge from dynamics of extra-dimensions which cannot be described in terms of any standard dynamics of matter. However,  they still scatter light non-trivially, and therefore they have gravitational lensing signatures that could be probed by future experiments.

Just like Schwarzschild black holes, topological solitons have an outer photon shell that surrounds the geometries and can be used to define their apparent size. Moreover,  this photon shell is unstable, and we have derived the associated Lyapunov and critical exponents.  For topological stars these exponents depend on its charge, but they are of the same order as Lyapunov and critical exponent of Schwarzschild black holes. However, we have shown that Schwarzschild topological solitons have an outermost photon shell with a Lyapunov exponent and a size remarkably similar to the Schwarzschild shadow. These similarities are remarkable, taking into account that the two metrics are very different.

We have pushed the comparison further by building our own ray-tracing code to study numerically the light-scattering properties of these solitons.
Through various illustrative plots,  we have demonstrated that topological stars are highly-coherent gravitational lenses that can be described as ``spacetime mirrors.''  Despite having a photon shell,  they do not have shadows: photons that ``go in'' generally bounce off without experiencing high redshifts or large time delays.  However, scattering in Schwarzschild topological solitons is much more complex: light that enters the photon shell of the solitons can have strong chaotic behavior, very high redshift and large elapsed time. These properties are expected to produce an effective scrambling behavior and a phenomenological horizon effect from regular gravitational structures.  All together, the Schwarzschild topological solitons have scattering properties very similar to Schwarzschild black holes.  The main difference will be a residual faded glow that emerges from inside the would-be shadow.

The present project has shown that topological solitons from quantum gravity are relevant to describe real-world physics and as macroscopic alternatives to black holes. This motivates further studies to better understand to what extent they are similar or different from black holes. The presence of smooth topological microstructures beyond the shadow should also have observational implications for the tidal stress that geodesics encounter when passing through the solitons \cite{Tyukov:2017uig,*Bena:2020iyw}, for quasinormal mode spectra \cite{Bena:2019azk,Bena:2020yii,QNMsTS}, and potentially also for gravitational-wave echoes \cite{Mayerson:2020tpn,Dimitrov:2020txx}.

Moreover, our computations have highlighted the presence of stable inner photon rings deep inside the topological microstructures. Even if they are not reachable from outside the outer photon shell, these trapped surfaces where matter can accumulate, radiate and interact with the soliton are associated with non-linear instabilities \cite{Cunha:2022gde,Keir:2014oka,*Cardoso:2014sna,*Cunha:2017qtt}.  In four-dimensional GR, they can lead to migration to non-ultracompact configurations or collapse to a black hole~\cite{Cunha:2022gde,Zhong:2022jke}. In string theory, additional quantum gravity degrees of freedom can induce geometric transitions and quantum tunnelings, so that these states may scramble to less coherent and more generic quantum states. Thus, the fate of the instability is still to some finite and non-singular states. While in the classical limit, it is expected that such generic states will be more and more indistinguishable from black holes, our work demonstrates that residual observable differences might remain. 

The accretion disk images we obtain in this paper were produced to highlight theoretical differences between black holes and topological solitons, not to  be experimentally realistic. In the future, we would like to provide a more experimentally-relevant analysis by modeling plasma orbiting topological solitons with full radiative transfer methods~\cite{Guo:2022nto,Vincent:2020dij}.

\acknowledgments 
\vspace{-0.1cm} 
We thank Iosif Bena,  Thomas Helfer,  Daniel Mayerson,  Masaki Shigemori, and Zackary White for interesting and helpful discussions.  The work of I.B. and P.H. is supported by NSF grant PHY-2112699.  The work of I.B. is also supported in part by the Simons Collaboration on Global Categorical Symmetries. E.B. is supported by NSF Grants No. AST-2006538, PHY-2207502, PHY-090003 and PHY-20043, and NASA Grants No. 19-ATP19-0051, 20-LPS20-0011 and 21-ATP21-0010.
This work was carried out at the Advanced Research Computing at Hopkins (ARCH) core facility (\url{rockfish.jhu.edu}), which is supported by the NSF Grant No. OAC-1920103.
The authors acknowledge the Texas Advanced Computing Center (TACC) at The University of Texas at Austin for providing {HPC, visualization, database, or grid} resources that have contributed to the research results reported within this paper~\cite{10.1145/3311790.3396656}. URL: \url{http://www.tacc.utexas.edu}.

\vspace{1.12cm}
\bibliography{microstates}

\clearpage \appendix

\end{document}